\documentclass{ifacconf}

\makeatletter
\let\old@ssect\@ssect 
\makeatother

\usepackage{graphicx}      
\usepackage{natbib}        
\usepackage{amsmath}        
\usepackage{amssymb}       
\usepackage{algorithm, algpseudocode}    
\usepackage{url}
\usepackage{hyperref}      
\usepackage{xcolor}        

\makeatletter
\def\@ssect#1#2#3#4#5#6{%
  \NR@gettitle{#6}
  \old@ssect{#1}{#2}{#3}{#4}{#5}{#6}
}
\makeatother
\begin{document}
\begin{frontmatter}

\title{Fault-Tolerant Control of Degrading Systems with On-Policy Reinforcement Learning} 


\author[First]{Ibrahim Ahmed} 
\author[First]{Marcos Quiñones-Grueiro} 
\author[First]{Gautam Biswas} 

\address[First]{Institute of Software Integrated Systems, Vanderbilt University, Nashville, TN, USA
   (e-mail: ibrahim.ahmed@vanderbilt.edu, marcosqg88@gmail.com, gautam.biswas@vanderbilt.edu).}

\begin{abstract}                
We propose a novel adaptive reinforcement learning control approach for fault tolerant control of degrading systems that is not preceded by a fault detection and diagnosis step. Therefore, \textit{a priori} knowledge of faults that may occur in the system is not required. The adaptive scheme combines online and offline learning of the on-policy control method to improve exploration and sample efficiency, while guaranteeing stable learning. The offline learning phase is performed using a data-driven model of the system, which is frequently updated to track the system's operating conditions. We conduct experiments on an aircraft fuel transfer system to demonstrate the effectiveness of our  approach.
\end{abstract}

\begin{keyword}
Fault tolerance, Reinforcement learning, Neural networks, Control system design, Machine learning 
\end{keyword}

\end{frontmatter}

\section{Introduction}

All systems are susceptible to degradation of components during system operation. With time, the degradation may increase to start impacting system operations. Fault-tolerant control (FTC) is defined as the ability of a system to continue operation under faulty conditions \citep{bla97}. The performance may be sub-optimal but the system degrades gracefully, and catastrophic failures are avoided. FTC systems are being applied to a wide range of industrial and transportation applications, primarily due to increased safety and reliability demands \citep{zhang2008}.

\cite{bla97} surveys the architecture of traditional FTC systems with two components: (1) detection and diagnosis systems, and (2) accommodation of faults to avoid complete system failures. Fail-safe systems are different from FTC systems because they introduce component redundancies to prevent failure conditions from occurring in the first place. Thus, FTC approaches work to mitigate the effects of faults after they occur, and require comparatively fewer hardware add-ons to the system.

\cite{noura2000fault} discusses FTC for sensor and actuator faults. For sensor faults, the controller does not respond, as the fault only impacts the perception of state. For actuator faults, the control law has to be modified. In both cases, accurate fault detection and diagnosis methods are required. In \citep{zhang2003fault, jiang2006accepting}, performance degraded reference models are used to develop a FTC system. Once a fault is diagnosed, a reference model matching the fault signature is employed for control. This can be applied to progressively degrading systems by generating reference models corresponding to various levels of degradation. This approach, however, requires \textit{a priori} knowledge of possible faults as well as a set of accurate system models for different degrading conditions.

Reinforcement learning (RL)-based control is an active area of research. RL relies on feedback from past experiences to generate future control actions in an automatic fashion. It has been applied to robotics \citep{kober2013reinforcement}, board games \citep{silver2018general}, and smart buildings \citep{naug2019online}. \cite{lewis2012reinforcement} discusses RL methods to design optimal and adaptive controllers. Adaptive controllers learn from a system model on the fly, but at the cost of true optimality in control. \cite{liu2016} focuses on speeding neural network training so the data-driven system model can be updated regularly in response to faults.  Responses to incipient and abrupt faults are demonstrated for a non-linear system. \cite{zhao2017} develops an RL-based fault-tolerant controller for actuator faults. A cost function is formulated by combining two terms: the cumulative sum of the effect of actuator faults on system's performance, and the effort of the control action over time. Overall, the goal is to minimize the cost function over the possible action space.

In our previous work, \cite{ahmed2018comparison}, we developed a RL-based fault-tolerant controller to accommodate abrupt faults. Faults were simulated on a fuel transfer system of a C-130 aircraft. A value iteration algorithm was used with a polynomial action-value function to derive a control policy. The controller showed better robustness to Gaussian sensor noise in comparison to model-predictive control (MPC). The proposed algorithm learned a policy offline with \textit{a priori} knowledge of faults applied to a system model. 

In this work, we extend our previous work and develop an online adaptive RL control algorithm for degrading systems with the goal 
of extending remaining useful life of a system, or providing sufficient time to continue degraded but safe operations till the system undergoes maintenance. 
Online RL-based control offers several advantages. By learning from past experience, it can adapt to changing system dynamics without changing design-time parameters, unlike classic control approaches like proportional integral derivative (PID) control. Moreover, compared to MPC schemes, a complex non-linear and non-convex optimization problem does not have to be solved online. This implies that the computational power required may be significantly less for RL-based control approaches during online operation.

The main contributions of this paper are the following: (1) the use of RL to learn a fault-tolerant controller that does not require \textit{a priori} knowledge of faults, or a fault detection and diagnosis step; and (2) the combination of online and offline policy learning to improve exploration and sample efficiency while guaranteeing stable learning. We demonstrate the effectiveness of the proposed approach by applying it to a complex non-linear system under different degrading conditions. We believe that this is a first application of an online/offline RL control scheme to FTC. 

The paper is organized as follows. First, the RL principles and the learning algorithm employed in this work are described. Then, the physics-based hybrid model of the operations for an aircraft fuel transfer system is presented.  We devise methods for seeding different incipient faults into the system. This is followed by a description of the FTC controller design using a mixed learning RL approach. The experiments and results are presented next to demonstrate empirically how the proposed controller adapts to different incipient fault scenarios. In the last section, we present the conclusions and directions for future work.
\section{Reinforcement Learning}


Reinforcement learning (RL) straddles the divide between supervised and unsupervised learning. In RL, unlike supervised learning, there are no solutions to learn from.  However, a feedback or \textit{reward} signal is provided for each (input, output) pair, such that a mapping can be created between inputs and outputs, from which the highest total reward can be learned over time.

An RL problem comprises of a controller (or an agent) and a system (or the environment). The controller can perceive the state of the system ($x_t \in X$). The controller acts on the system by generating an action ($u_t$). The system responds by evolving its state ($x_{t+1}$) and generating a reward (or feedback) signal ($r_{t+1}$). This cycle of perception, action, and reward completes a single interaction which is typically called experience. An episode in RL lasts from an initial state and ends when some terminal condition is reached.

The goal of RL is to derive an \textit{optimal policy} $\pi^*$. A policy function maps a state to an action $\pi: x \rightarrow u$. An optimal action is one such that the total expected future rewards are maximized. The total expected future reward from a state under an optimal policy is defined as its true value $V^*(x)$. Delayed rewards may be discounted by a factor $\gamma \in [0, 1]$ to emphasize immediacy.

\begin{equation} \label{eq:value}
V^*(x_t) = \max_{u} ( \mathbb{E}[r + \gamma V^*(x_{t+1})] )
\end{equation}

The return of a state $x_0$ under any policy $R^\pi(x_0)$ is the cumulative discounted reward by following that policy from that state. The expected return under a policy is the estimated value of a state. 

\begin{equation}\label{eq:return}
    R^\pi(x_t) = \sum_t (\gamma^t r_{t+1}),\;\;
    V^\pi(x_t) = \mathbb{E}[R^\pi(x_t)]
\end{equation}

In policy gradient algorithms \citep{sutton2000policy}, the policy function is parameterized $\pi_\theta(x): x \rightarrow u$ and learned through gradient ascent on the expected returns. The magnitude of change in policy parameters from previous version $\theta_k$ to the next $\theta_{k+1}$ is governed by the step size $\alpha$.

\begin{equation} \label{eq:policygradient}
    \theta_{k+1} = \theta_{k} + \alpha \nabla_\theta \mathbb{E}[V^{\pi_\theta}(x)]
\end{equation}

\textit{Proximal Policy Optimziation (PPO)} 

PPO is a policy gradient algorithm that can be used for systems with either discrete or continuous action spaces \citep{schulman2017proximal}. PPO is an actor-critic algorithm, where the policy function $\pi_\theta(x)$ and value function $V_\theta(x)$ are both parametrized. By modeling $V^\pi(x)$, the algorithm does not need to play out multiple trajectories of a policy to estimate a state's value. PPO can be classified as an on-policy reinforcement learning method because the updates of the policy are based on the experiences collected under the up-to-date policy. Learning through policy gradient methods is challenging because they are sensitive to the choice of update step size, i.e. too small a step makes convergence extremely slow (requiring thousands of interactions with the environment), and too big  a step makes them prone to divergence. Thus, researchers have sought strategies to make the step size sensitive to the learning process. This is the idea behind PPO. PPO differentiates itself from other on-policy, policy gradient algorithms like Trust Region Policy Optimization (TRPO) by having a computationally simpler estimation of the step size.

PPO includes mechanisms to constrain the update step size while using first-order optimizers, like the gradient descent method, by formulating the following objective function:

\begin{equation} \label{eq:ppo}
\begin{aligned}
    \theta_{k+1} &= \arg \max_{\theta} \mathbb{E}_{x,u \sim \pi_{\theta_k}}[L(x,u,\theta,\theta_k)] \; \mathrm{s.t.}\\
    L(x,u,\theta,\theta_k) &= \min\Bigg[ \\ 
    (R^{\pi_{\theta_k}}(x) - &V_{\theta_k}(x)) \cdot \frac{\pi_{\theta}(u|x)}{\pi_{\theta_k}(u|x)} , \\
    (R^{\pi_{\theta_k}}(x) - &V_{\theta_k}(x)) \cdot clip \bigg(\frac{\pi_{\theta}(u|x)}{\pi_{\theta_k}(u|x)},1-\epsilon,1+\epsilon)\bigg) \Bigg],\\
\end{aligned}
\end{equation}
where $\epsilon$, called the penalty coefficient, is the hyper-parameter that controls how much the new policy is allowed to differ from the old one.  If the probability ratio between the new policy and the old policy falls outside the range $[(1-\epsilon),(1+\epsilon)]$ then the clip function is applied to cut-off the update step size. 

\textit{Off-policy and On-policy updates}

For learning a control policy, the choice of off-policy versus on-policy updates is related to the bias-variance trade-off \citep{Sigaud2019}. Off-policy RL algorithms, such as those based on Q-learning, perform policy updates by reusing samples from previously collected experiences generated by a different policy, which are typically stored in a replay buffer. Off-policy updates allow for greater exploration of states, but such updates introduce a bias, especially for time-varying systems, where the experiences used for the updates may have been generated under a different operating condition for the system. The bias in the update step can make the policy sub-optimal or even divergent. Conversely, on-policy RL algorithms allow for more stable learning. However, they are less sample efficient since more interactions with the environment are required to converge.

\textit{Offline and online updates}

In this paper we propose a mixed learning scheme. We employ a data-driven model of the system. The same on-policy learning algorithm (PPO) is used to improve the current policy but the interactions are performed offline with the system's model periodically to improve the exploration process. The model is updated (relearned) at frequent periodic intervals to reflect changes in the system as it degrades over time. Employing a model as an offline environment allows the controller more interactions to experience system behavior and feedback. Thus, it is more sample efficient.
\section{Fuel Transfer System}

\begin{figure}[ht]
\begin{center}
\includegraphics[width=6cm]{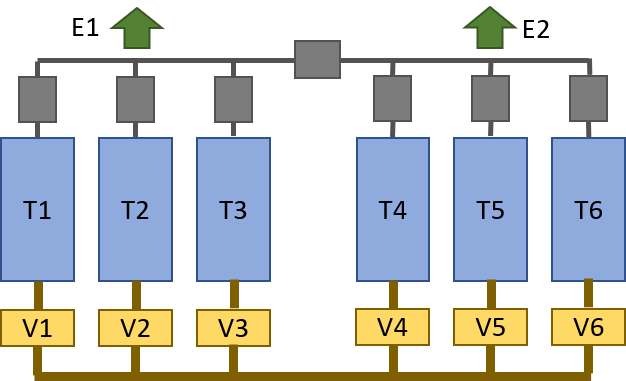}    
\caption{Simplified fuel system schematics. The controller manages valves and can observe fuel tank levels. Net outflow to engines via pumps is controlled independently. Pumps drain tanks innermost first.}
\label{fig:fuelsystem}
\end{center}
\end{figure}

This section describes the operations of a simplified fuel transfer system for C-130 cargo planes (see  Fig.~\ref{fig:fuelsystem}) that we use for our case study. The system comprises of six fuel tanks, three on either side of the fuselage. The geometry of tanks is assumed to be identical with a cross section $\vec{c}$. Under normal operations the fuel in the tanks on each side are pumped into the corresponding engines. All six tanks are also connected to each other through a common fuel conduit, and flows between the tanks are controlled by valves. The fuel exchange between tanks, when the corresponding valves are open, is governed by the pressure differential between the tanks. An optimal distribution of fuel maintains the centre of gravity in the aircraft, while keeping the mass distributed across the fuel tanks for stability.

During operation, the pumps maintain a flow rate from the active tanks based on the fuel demand of the engines. The transfer paradigm first delivers fuel from the two innermost tanks to the engines, and then switches to the outer tanks.

Valves are only opened to transfer fuel to the engines or to redistribute fuel between tanks through the shared conduit. The flow rate from and into tanks is governed by the gravitational potential of fuel and the resistance of valves. When valves are opened, the tank with the highest potential becomes the flow source, while the remainder of tanks with valves open become sinks. The flow rate into sinks is proportional to the potential difference with the source and inverse of the valve resistance along the flow path. The flow rate from each source tank is proportional to the inverse of its valve resistance.

The state $\vec{x}$ of the system is described by six fuel tank levels. The action space $\vec{u}$ is six-dimensional and binary, corresponding the status of each valve. A ``0" means closed and a ``1" means open.




Multiple degradation points are present in the fuel transfer system. Valve resistances $\vec{r_v}$ in tanks can increase due to deposition of residue over time. Pumping capacity of engines $\vec{p}$ can deteriorate over time due to wear and tear of the pumps. Similarly fuel demand $\vec{d}$ of the engines can increase unevenly as the engines age and become more inefficient. Unless the nominal control paradigm is modified, this will also cause fuel imbalance.

In this paper, as an approximation, system degradation is modeled as a linear function of time. Degradation is modeled separately for engines and tanks. The degradation factors for engines and tanks ($D_E, D_T$) are the time it takes for a particular system parameter to increase by a 100\% ($\vec{r_v}, \vec{d}$) or decrease by a 100\% ($\vec{p}$) from its initial nominal value.

\section{Approach}

\begin{figure}[ht]
\begin{center}
\includegraphics[width=6cm]{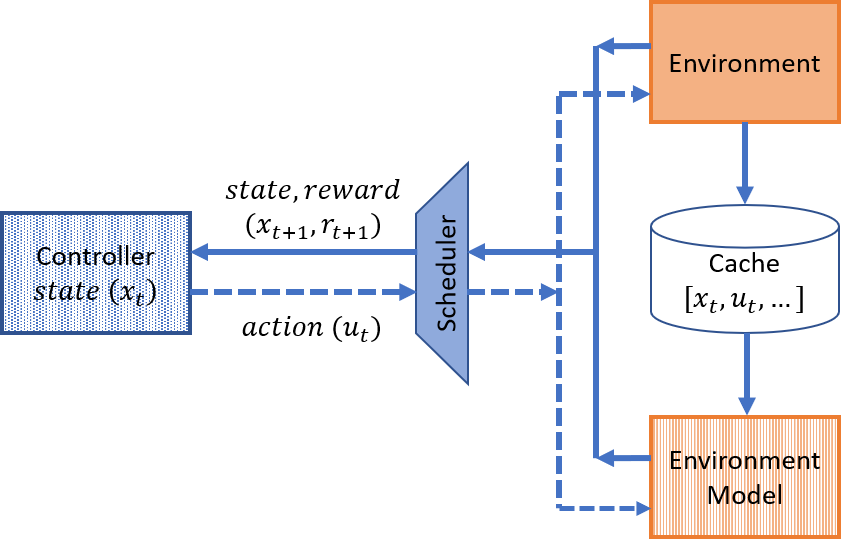}    
\caption{An overview of the proposed approach. The controller learns from the system and its data-driven model model to improve sample efficiency. Online control may proceed while the controller trains offline on the model in the background.}
\label{fig:overview}
\end{center}
\end{figure}

The control loop is discussed in this section. This paper proposes an on-policy reinforcement learning-based control loop with a mixed learning scheme. The control policy is updated based on interactions with the real system (online). Intermittent periods of offline updates are also carried based on interactions with a data-driven model of the degrading system. The model is periodically updated to account for changing conditions in the system due to degradation. The controller logic is presented in Algorithm \ref{alg:ctrl}, with the overview illustrated in figure \ref{fig:overview}.

During operation, the controller for the fuel transfer system follows the derived policy function. Each interaction is cached. After $t_{update}$ interactions, the policy function is updated online. After an interval $t_{online}$ interactions, a data-driven model of the system is learned from cached experience. Offline, the controller updates its policy using the model for $t_{offline}$ interactions.

Using an offline learning phase addresses the problem of sample inefficiency of on-policy reinforcement learning algorithms. This is due to two reasons: (1) during operation, the size of experience of on-policy controllers is limited by the frequency of interactions with the system; and (2) on-policy controllers, by design, are restricted to applying a single policy during learning and operation. Therefore, they cannot benefit from sampling alternative behaviours.

Given the availability of an accurate model of the system, the controller is free to interact at a frequency of its choosing, bounded by two criteria: (1) that a dynamic model of the system is sufficiently accurate to derive an ``optimal'' policy $-$ degradation of components in the system may cause the model to become inaccurate, so the model is re-learnt at frequent periodic intervals. And (2) computational resources.

A fully connected neural network is employed as a system model. The next state of the system solely depends on its current state and action taken. The sampling frequency of the data is high enough to capture system dynamics and there are no time delays. Furthermore, the state and action spaces are compact. This satisfies the requirements of the universal approximation theorem for neural networks (\cite{cybenko1989approximation}).

\begin{algorithm}[ht]
\caption{Proposed reinforcement learning-based control loop.}
\label{alg:ctrl}
\begin{algorithmic}[1]
\Require $t_{online}$ \Comment{Duration of online control}
\Require $t_{offline}$ \Comment{Duration of offline learning}
\Require tanks, engines degradation factors
\Require offline $\in {\texttt{True, False}}$
\State initialize policy function $\pi$, system model
\Loop
\State degrade system
\For{$t=0$ to $t_{online}$}
    \State Apply control action $u=\pi(x)$ from state
    \State Cache experience (states, actions, rewards)
    \If{$\mod{t, t_{update}}=0$}
        \State Update policy $\pi$ given system
    \EndIf
\EndFor
\If {offline=\texttt{True}}
    \State Update system model from cache
    \State Clear cache
    \For{$t=0$ to $t_{offline}$}
        \State Apply control action $u=\pi(x)$
        \If{$\mod{t, t_{update}}=0$}
            \State update policy $\pi$ given system model
        \EndIf
    \EndFor
\EndIf
\EndLoop
\end{algorithmic}
\end{algorithm}

\section{Experiments}

\subsection{Environment Formulation}

The fuel transfer model is considered as the controller's environment. An episode begins when the fuel tanks are at capacity, i.e., at the beginning of an aircraft flight, with the assumption that the system is operating nominally. The episode ends when total fuel in the tanks is less than the total demand from engines, implying that the pilot would have maneuvered the aircraft to land just before this situation occurred, which would cause the aircraft to crash. The environment is run for 10 intervals of $t_{online}=512$ steps each. A single interval contains multiple episodes. After each interval, the system is degraded. For offline learning, the controller interacts with the model for an interval of $t_{offline}=2048$ steps.

The objective of the controller is to maintain aircraft stability (i.e., center of gravity) even under degrading fault conditions in the aircraft. The reward function, defined in equation \ref{eq:reward}, is comprised of three variables: (1) centre of gravity ($r_{cg}$), (2) variance of fuel distribution $r_{var}$, and (3) the proportion of open valves ($r_u$). A centre of gravity close to the longitudinal axis of the aircraft contributes to lateral stability. Similarly, concentration of mass about the wing tips for forward-swept wings makes the plane more stable (\cite{goraj2005aircraft}). Finally, it is desirable to have a minimal number of valves open for fuel transfer so that fluid mass does not shift when a plane maneuvers in air. Fuel tanks were assumed to be equally spaced, unit distance apart from each other about the fuselage axis. The moment arms of fuel tanks are given by $\vec{m} = (-3, -2, -1, 1, 2, 3)$.

\begin{figure}[ht]
\begin{center}
\includegraphics[width=6cm]{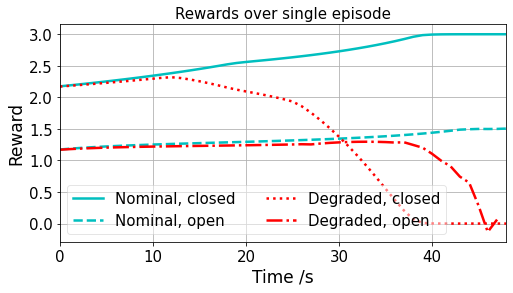}    
\caption{Comparison of rewards per step over an episode between nominal and degraded operation with open and closed valves. The left-most engine is degraded. This causes fuel on left tanks to drain faster. As the centre of gravity shifts to the right, the contribution of variance is scaled down.}
\label{fig:rewardsepisodecomparison}
\end{center}
\end{figure}

The reward function (equation \ref{eq:reward}) prioritizes a neutral centre of mass. $\odot$ denotes the Hadamard product. Only when centre of mass is close to the axis is the variance of fuel distribution rewarded. Figure \ref{fig:rewardsepisodecomparison} illustrates how reward changes over an episode.

\begin{equation}\label{eq:reward}
\begin{aligned}
r_{cg}  &= \sum(\vec{x} \odot \vec{c} \odot \vec{m}) / (\vec{x} \cdot \vec{c}) \\
r_{var} &= \mathrm{var}_{\vec{x} \odot \vec{c}}(\vec{x} \odot \vec{c} \odot \vec{m}) \\
r_{u}   &= \bar{u} \\
r       &= \left(1 - \frac{\lvert r_{cg} \rvert}{\max \vec{m}} \right) \cdot r_{var} - r_{u}
\end{aligned}
\end{equation}

\subsection{Data-driven System Model}

A hyperparameter grid search was conducted to obtain the most accurate model architecture for the fault-free fuel transfer system. Training data was generated from the model for 50 episodes. Random actions were chosen, which persisted for at most half an episode length. We noted that increasing the width of networks improved performance more than increasing depth. The best performance model had 2 hidden layers of 64 units each with Rectified Linear Unit (ReLU) activations. It was trained with an Adam optimizer learning algorithm with early stopping and a learning rate of $10^{-3}$. A coefficient of determination of $0.996$ was achieved with cross-validated data. Models used in the following experiments were initialized to the learned parameters from this model.

\subsection{A single fault: engine degradation}

For the trial, a degrading fault was introduced in the left-most engine, ($E_1$) with degradation factor $\vec{D_E}[E_1] = 20$. The system degraded linearly with each interval.

\begin{figure}[ht]
\begin{center}
\includegraphics[width=6cm]{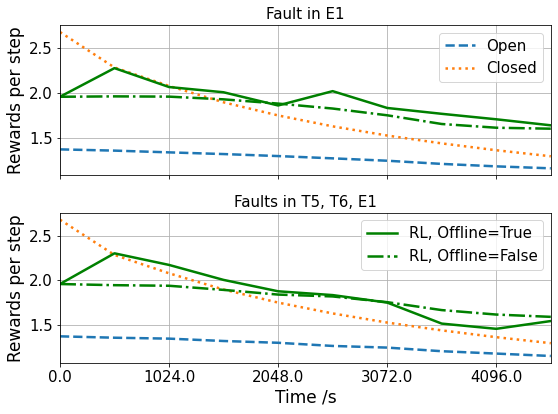}    
\caption{Average rewards per time step for each online interval as system progressively degrades under a single (top) and multiple (bottom) faults.}
\label{fig:rewardssingle}
\end{center}
\end{figure}

Figure \ref{fig:rewardssingle} illustrates controller performance under progressing degradation over multiple intervals. When all valves are closed, the fuel distribution is no longer symmetric and the centre of gravity shifts to the right. Average rewards per step rapidly decline with increasing degradation. However, fuel does not freely flow between tanks, so the variance remains large, hence the high initial rewards when faults are minor. When all valves are open, fuel is able to freely move between tanks. The distribution of fuel is kept symmetric about the longitudinal axis. This also means that fuel rushes to median tanks as they drain first. This is penalized by the variance term $r_{var}$ in the reward.

\begin{figure}[ht]
\begin{center}
\includegraphics[width=6cm]{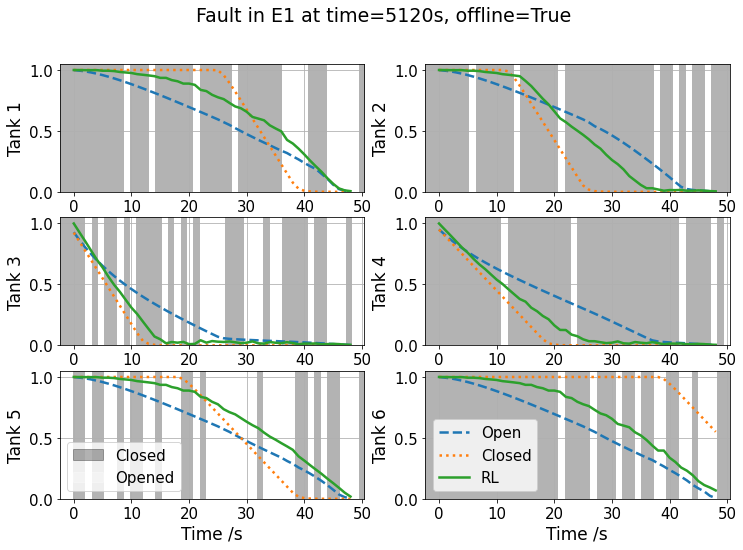}    
\caption{Control of fuel tanks when the leftmost engine is degraded.}
\label{fig:controlsingle}
\end{center}
\end{figure}

Figure \ref{fig:controlsingle} shows control actions over one episode at the highest degradation. The RL agent maintains a better balance than when all valves are open or closed at high levels of degradation. Valves open for tanks on the opposite side to equalize fuel distribution, as incentivized by $r_{cg}$. Of note is how tank 5's valve is primarily opened up to transfer fuel to tank 3 (Open valves are illustrated by the white spaces in the tank level plots in figure \ref{fig:controlsingle}). This allows tank 6 to maintain its fuel mass and earn the reward $r_{var}$ for a high distribution variance.

\subsection{Multiple faults: Engine and tank degradation}

Another trial was carried out where the leftmost engine was degraded the same as before. Additionally the valves and resistances for tanks 5 and 6 were degraded with a factor $\vec{D_T}[T_5, T_6] = 20$. Figure \ref{fig:rewardssingle} shows that the difference in performance between online-only and online+offline control schemes is less noticeable under multiple faults, but still outperforms baseline.

\begin{figure}[ht]
\begin{center}
\includegraphics[width=6cm]{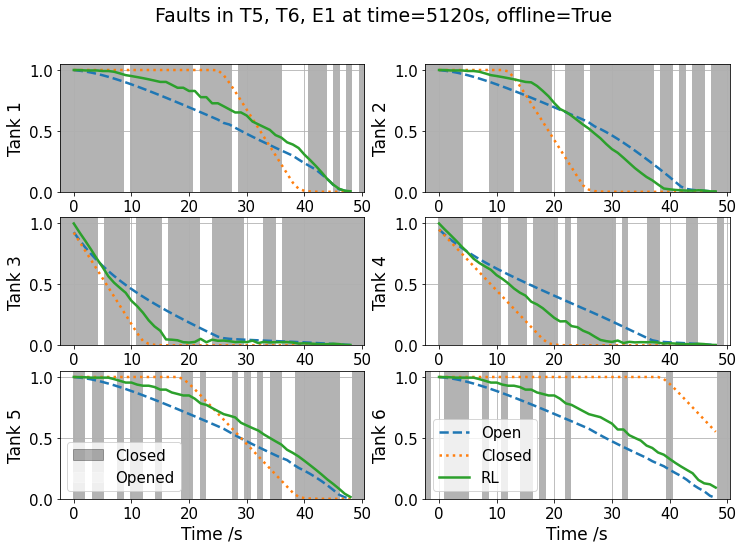}    
\caption{Control of fuel tanks when the leftmost engine and two right tanks are degraded.}
\label{fig:controlmultiple}
\end{center}
\end{figure}

\subsection{Aggregate performance}

\begin{figure}[ht]
\begin{center}
\includegraphics[width=6cm]{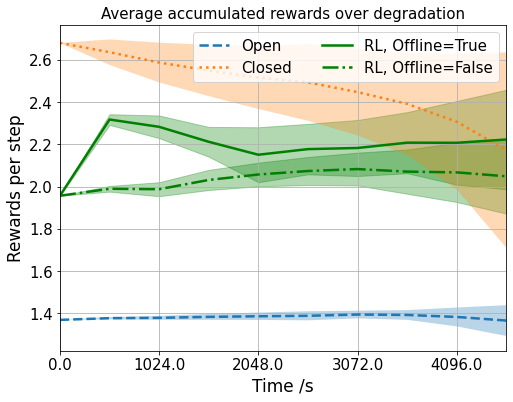}    
\caption{Average rewards per time step for each interval of $t_{online}$ steps as system progressively degrades under multiple faults. A big contribution to offsets between rewards is $r_u$ which rewards closed valves.}
\label{fig:rewardsagg}
\end{center}
\end{figure}

20 trials were carried out and the results averaged. For each trial, a valve and and an engine pump (chosen randomly) were assigned a degradation factor between $[10, 30]$. Figure \ref{fig:rewardsagg} shows aggregate performance. Rewards earned by RL control do not show deterioration as degradation increases. RL control with offline learning consistently outperforms other approaches.

\section{Conclusion}

The fault-tolerant control problem has been applied to a wide range of transportation and industrial applications \citep{zhang2008}.  This paper discusses an on-policy reinforcement learning-based fault-tolerant control scheme for incipient faults that does not require the fault detection and isolation step. Control is tested on a degrading model of an aircraft fuel transfer system. The controller is incentivized to maintain a high fuel distribution variance, centre of gravity close to longitudinal axis of the aircraft, and a minimal number of open valves. Under these constraints, RL control outperforms the baseline case when all valves are kept closed. On average, RL-based control shows a slower degradation in performance than the baseline cases (all valves closed or all valves open).

Previous methods to fault tolerance require a fault detection and isolation or performance monitoring to trigger the adaptation schemes  \citep{li2019performance, zhang2008}. The proposed approach does not rely on fault detection and isolation to learn control policies under degraded conditions. As the system degrades, the on-policy algorithm is able to optimize actions based on changing reward signals. In contrast to model-predictive control, an optimization problem is not solved at each time step. This points towards FTC applications where memory (for storing experiences) is abundant and the limiting factor is computational power.

There are several directions for future research. Fault-detection mechanisms can be explored, such that offline control re-learning is scheduled aperiodically whenever faults are detected. The control scheme can also be applied to the case of abrupt faults. Finally, a thorough study is merited into the choice of reward functions, model architecture, and convergence guarantees for a control policy across a general class of non-linear hybrid systems.\footnote{Source code for this work can be found at \url{https://git.isis.vanderbilt.edu/ahmedi/airplanefaulttolerance/-/tree/ifac2020}}


\bibliography{references}


\end{document}